\documentclass[12pt,preprint,number,sort&compress,lontitle,times]{elsarticle}
\usepackage{natbib}
\usepackage{graphicx}
\usepackage{epsfig}
\usepackage{lineno}
\usepackage{multirow}
\usepackage{fourier}
\usepackage{color}
\usepackage{appendix}
\usepackage[english]{babel}

\begin{document}

\begin{frontmatter}
\title{An Estimate of the Spectral Intensity \\
Expected from the Molecular Bremsstrahlung Radiation \\
in Extensive Air Showers}
\author{I. Al Samarai$^{1}$, O. Deligny$^{1}$, D. Lebrun$^2$, A. Letessier-Selvon$^3$, F. Salamida$^{1}$ \\
$^{1}$ Institut de Physique Nucl\'{e}aire d'Orsay, \\
CNRS/IN2P3 \& Universit\'{e} Paris Sud, Orsay, France\\
$^2$ Laboratoire de Physique Subatomique et Corpusculaire, \\
CNRS/IN2P3 \& Universit\'{e} Joseph Fourier, Grenoble, France\\
$^3$ Laboratoire de Physique Nucl\'{e}aire et des Hautes Energies, \\
CNRS/IN2P3 \& Universit\'{e} Pierre et Marie Curie, Paris, France
}

\begin{abstract}
A detection technique of ultra-high energy cosmic rays, complementary to the fluorescence technique, 
would be the use of the molecular Bremsstrahlung radiation emitted by low-energy electrons left after
the passage of the showers in the atmosphere. The emission mechanism is expected from quasi-elastic
collisions of electrons produced in the shower by the ionisation of the molecules in the atmosphere. \\
In this article, a detailed calculation of the spectral intensity of photons at ground level originating from the
transitions between unquantised energy states of free ionisation electrons is presented. In the absence
of absorption of the emitted photons in the plasma, the obtained spectral intensity is shown to be
$\simeq 4.0~10^{-26}$~W~m$^{-2}$~Hz$^{-1}$ at 10~km from the shower core for a vertical shower induced 
by a proton of $10^{17.5}$~eV.
\end{abstract}
\end{frontmatter}


\section{Introduction}

The origin and nature of ultra-high energy cosmic rays still remain to be elucidated despite the 
recent progresses provided by the data collected at the Pierre Auger Observatory and the Telescope 
Array~\cite{KHK-PT}. This is due to the extremely low intensity of particles at these energies. 
As of today, the most direct way to infer the nature of the particles at ultra-high energies relies on 
the observation of the shower longitudinal profile to measure its maximum of development.
The use of telescopes detecting the nitrogen fluorescence light emitted after the passage
of the electromagnetic cascade is a well-suited technique to achieve such measurements.
Moreover, these fluorescence telescopes provide a good calorimetric estimate of the energy of 
the showers, which is preferable to detectors requiring external information to calibrate the energy 
estimator of the showers. However, this technique can only be used on moonless nights, resulting 
in a 10\% duty cycle. Together with the low intensity of particles, this makes the study of the cosmic 
ray composition above few tens of EeV very challenging.

Triggered by microwave emission measurements in laboratory~\cite{Gorham}, new telescope 
techniques based on the detection of the microwave emission in the GHz C-band (3.4-4.2 GHz) 
have been developed at the Pierre Auger Observatory~\cite{Gaior}. These techniques aim at providing 
measurements of the electromagnetic content of the cascade with quality comparable to the fluorescence 
detectors but with a 100\% duty cycle. Molecular Bremsstrahlung radiation in the GHz band provides an 
interesting mechanism to detect ultra-high energy cosmic rays due to the expected isotropic and 
unpolarised radiation. This feature would allow for the possibility of performing shower calorimetry
in the same spirit as the fluorescence technique does, by mapping the ionisation content along the
showers through the intensity of the microwave signals detected at ground level. 

Attempts to estimate the spectral intensity expected from the molecular Brems\-stra\-hlung radiation in beam
experiments~\cite{Gorham} or in extensive air showers~\cite{KITicrc2013} have been performed, based 
on general frameworks pertaining to radiative processes in plasmas. In these works, the sources
of the emission are the low-energy electrons left along the shower track after the passage of high-energy 
electrons of the cascade propagating in the atmosphere. The different energy distributions of the
ionisation electrons are considered as static during the time the electrons can emit. These approaches
resulted in a free-parametric~\cite{Gorham} or in a very low expectation~\cite{KITicrc2013} for the
signal power that could be observed at the ground level. 

In this paper, the approach adopted is based on the computation of the spectral power per volume unit, 
which is shown to be the natural quantity to estimate the spectral intensity at any reference point in space
and time. It is derived from the collision rate of ionisation electrons leading to the production of photons through 
free-free transitions. Moreover, the ionisation electrons are tracked from their production to their disappearance 
by accounting for all interactions affecting their energy distribution with time, as detailed in section~\ref{electrons}. 
In turn, these electrons can produce their own emission, such as Bremsstrahlung emission. The expected 
spectral intensity at ground level of such an emission is the object of section~\ref{mbr}. Possible attenuation 
or suppression effects are studied in section~\ref{attenuation_effects}. Finally, the results obtained in this
study are illustrated in section~\ref{discussion} on a toy reference shower. From these results, the 
perspectives of detection of ultra-high energy cosmic rays by making use of molecular Bremsstrahlung 
radiation are discussed.

\section{Ionisation Electrons along the Shower Track}
\label{electrons}

\subsection{A Crude Model of Vertical Air Showers}

In this work, an extensive air shower is considered as a thin plane front of high energy charged particles 
propagating in the atmosphere at the speed of light $c$. For a given primary type and a given energy $E$,
the longitudinal development of the electromagnetic cascade depends only on the \textit{cumulated slant depth}
$X$ expressed as the ratio between the vertical thickness of the atmosphere $X_{\mathrm{vert}}$ (1000 g~cm$^{-2}$
at sea level) and the cosine of the zenith angle of the shower. After the succession of a few initial steps in the 
cascade, all showers can be described by reproducible macroscopic states. In particular, the shape of the 
showers is universal except for a translation depending logarithmically on $E$ and a global factor roughly
linear in $E$. In this way, for any given slant depth $X$ or equivalently any altitude $a$, the total number of 
primary $e^+/e^-$ particles, $N_{e,p}$, can be adequately parameterised by the Gaisser-Hillas function 
as~\cite{GaisserHillas}:
\begin{equation}
N_{e,p}(a)=N_{\mathrm{max}}\bigg(\frac{X(a)-X_0}{X_{\mathrm{max}}-X_0}\bigg)^{\frac{X_{\mathrm{max}}-X_0}{\lambda}}\exp{\bigg(\frac{X_{\mathrm{max}}-X(a)}{\lambda}\bigg)},
\end{equation}
with $X(a)$ the depth corresponding to the altitude $a$, $X_0$ the depth of the first interaction,
$X_{\mathrm{max}}$ the depth of shower maximum, $N_{\mathrm{max}}$ the number of particles
observed at $X_{\mathrm{max}}$, and $\lambda$ a parameter describing the attenuation of the shower.

On the other hand, high energy particles constituting the \textit{core} of the shower are collimated along the
initial shower axis. The lateral extension of the core depends on the mean free path and can be expressed
in terms of the \textit{Moliere radius} $R_M$ such that 90\% of the energy is contained within a distance $r$ from 
the axis such as $r<R_M$. Motivated by general arguments to describe the electromagnetic cascade of showers, 
the NKG lateral distribution function denoted hereafter by $g(r,a)$ is known to reproduce reasonably well the
observations~\cite{NKG}:
\begin{equation}
g(r,a)=C(s(a))~R_M^{-2}~\left(\frac{r}{R_M}\right)^{s(a)-2}\left(1+\frac{r}{R_M}\right)^{s(a)-4.5}.
\end{equation}
Here, $s(a)$ stands for the age parameter at altitude $a$ defined as $s(a)=3X(a)/(X(a)+2X_{\mathrm{max}})$,
and $C(s)$ is a normalisation factor. 

\begin{figure}[!t]
\centering
\includegraphics[width=12cm]{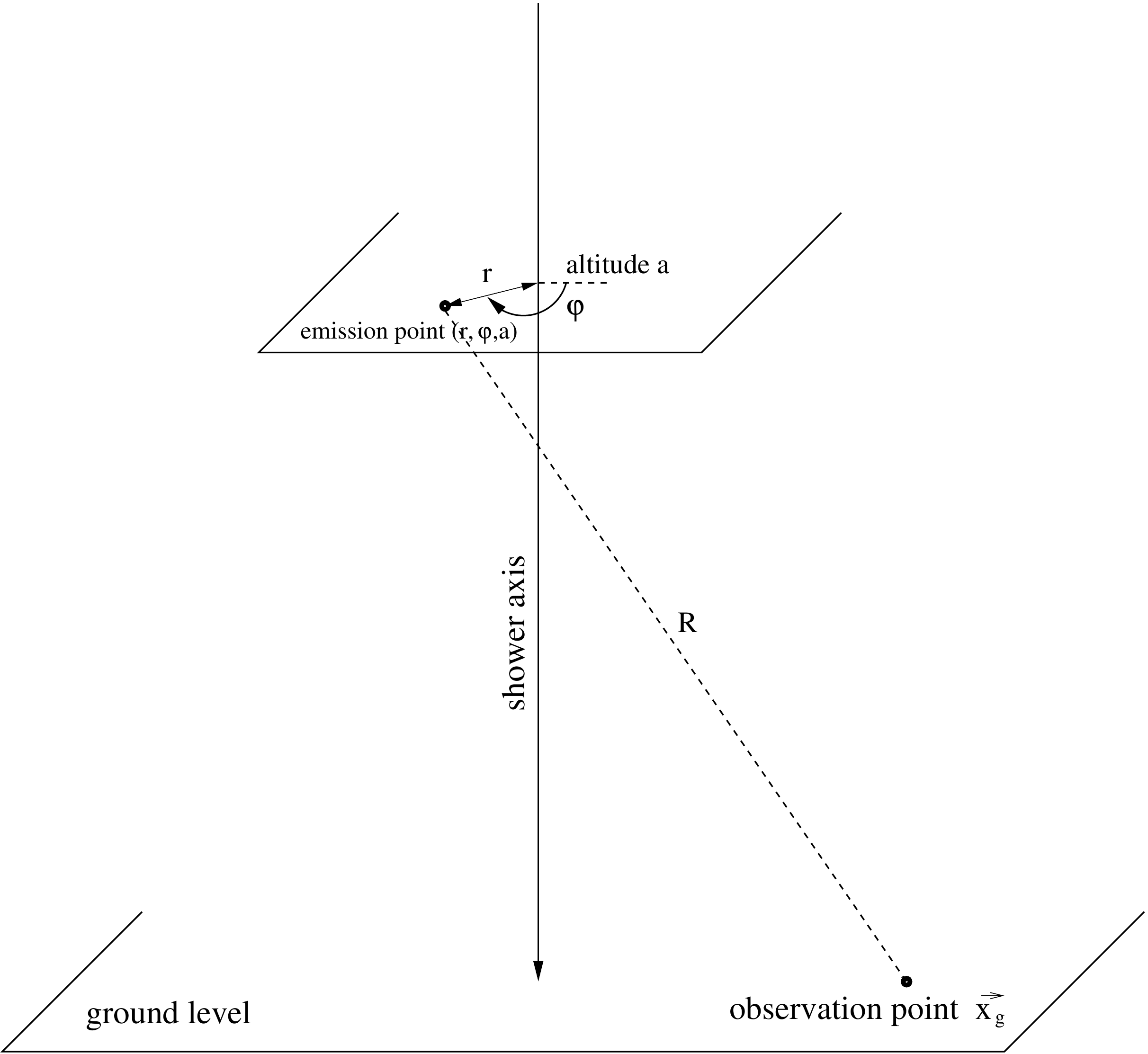}
\caption{\small{Geometry of a vertical shower used throughout the paper.}}
\label{fig:shower}
\end{figure}

The number of primary $e^+/e^-$ per unit surface, $n_{e,p}(r,a)$, is then simply obtained by
folding the longitudinal profile to the normalised lateral one. For a vertical shower whose geometry is depicted 
in figure~\ref{fig:shower}, $n_{e,p}(r,a)$ reads as:
\begin{equation}
n_{e,p}(r,a)=N_{e,p}(a)~\frac{g(r,a)}{2\pi\displaystyle\int \mathrm{d}r~r~g(r,a)}.
\end{equation}

Noticeably, this description is only a crude model of an extensive air shower. This shall allow us, however, to derive 
in the following a realistic number of ionisation electrons left along the shower track and thus to estimate relevant 
orders of magnitude for the spectral intensities (in W~m$^{-2}$~Hz$^{-1}$) that can be expected from molecular 
Brems\-strah\-lung radiation by these ionisation electrons. 

To facilitate comparisons of the results obtained in this study with the values reported in~\cite{Gorham} 
and~\cite{KITicrc2013}, the parameters of both the Gaisser/Hillas and the NKG functions are tuned to apply to 
vertical proton showers with primary energy $E=10^{17.5}~$eV.

\subsection{Production of Ionisation Electrons along the Shower Track}

Through the passage of charged particles in the atmosphere, the energy of an extensive air shower is deposited
mainly through the ionisation process. The resulting numerous ionisation electrons can, in turn, produce their own 
emission such as continuum Bremsstrahlung emission through quasi-elastic scattering with molecular nitrogen and, 
to smaller extent, oxygen. To evaluate the spectral intensity of this radiation, we start by deriving below the flux of 
secondary ionisation electrons created by the development of the shower.

For one single primary electron travelling over an infinitesimal distance $\mathrm{d}a$, and for a mass density 
$\rho_m(a)$ of molecular nitrogen or oxygen, the average number of ionisation electrons per unit length and per 
kinetic energy band reads as:
\begin{equation}
\label{d2N_dadT}
\frac{\mathrm{d}^2N_{e,i}}{\mathrm{d}a~\mathrm{d}T_e}(a,T_e)=\rho_m(a)~f_0(T_e)\left\langle\frac{\mathrm{d}E}{\mathrm{d}X}\right\rangle~\frac{1}{I_0+\left\langle T_e\right\rangle},
\end{equation}
with $I_0$ the ionisation potential to create an electron-ion pair in air ($I_0=15.6~$eV for N$_2$ and $I_0=12.1~$eV 
for O$_2$ molecules). The bracketed expression $\left\langle\frac{\mathrm{d}E}{\mathrm{d}X}\right\rangle$ stands for 
the mean energy loss of primary electrons per grammage unit. This energy loss is due quasi-exclusively to ionisation 
and is almost independent of the primary electron energy over a range of few tens of MeV, typical of the primary 
electrons energy in the cascade. The distribution in kinetic energy of the resulting ionisation electrons is described here 
by the normalised function $f_0(T_e)$. This distribution has been experimentally determined and accurately parameterised
for primary electrons with kinetic energies $T_e^p$ up to several keV~\cite{Opal}. For higher kinetic energies $T_e^p$,
relativistic effects as well as indistinguishability between primary and secondary electrons have been shown to modify
the low-energy behaviour~\cite{Arqueros}. To account for these effects, we adopt the analytical expression provided
in~\cite{Arqueros2}:
\begin{equation}
\label{eqn:f0}
f_0(T_e)=K\frac{1+C\exp{(-T_e/T_k)}}{T_e^2+\overline{T}^2},
\end{equation}
where $T_e$ ranges from 0 to $T_e^{\mathrm{max}}=(T_e^p-I_0)/2$ due to the indistinguishability between primary
and secondary electrons, the constant $C$ is determined in the same way as in~\cite{Opal} so that 
$\int \mathrm{d}T_e~C/(T_e^2+\overline{T}^2)$ reproduces the total ionisation cross section, 
$T_k=77~$eV is a parameter acting as the boundary between close and distant collisions,
the constant $K$ is tuned to guarantee $\int \mathrm{d}T_e f_0(T_e)=1$, 
and $\overline{T}=13.0~(17.4)~$eV for nitrogen (oxygen). In the energy range of interest, this expression leads to 
$\left\langle T_e\right\rangle\simeq 40~$eV, in agreement with the well-known stopping power. 
The instantaneous number of ionisation electrons per unit volume and per kinetic energy band is then obtained by 
coupling equation~\ref{d2N_dadT} to the number of primary charged particles per surface unit:
\begin{equation}
n_{e,i}^0(r,a,T_e)=\frac{f_0(T_e)}{I_0+\left\langle T_e\right\rangle}~\left\langle\frac{\mathrm{d}E}{\mathrm{d}X}\right\rangle~\rho_m(a)~n_{e,p}(r,a).
\end{equation}

Of relevant importance for the following is the \textit{flux} $\phi_{e,i}^0(r,a,T_e)$
of secondary electrons per kinetic energy band. For any surface element $\mathrm{d}S$,
and considering a coordinate system with the zenith angle defined along the axis
perpendicular to the surface $\mathrm{d}S$, the total number of electrons $N_{e,i}(r,a,T_e,\chi,\psi)$ 
(per kinetic energy band) crossing $\mathrm{d}S$ during a short time interval $dt$ under zenith and 
azimuth incidence angles $\chi$ and $\psi$ is~:
\begin{equation}
\label{eqn:Nei}
N_{e,i}^0(r,a,T_e,\chi,\psi)=c\beta(T_e)\mathrm{d}t~\int \mathrm{d}\psi\sin{\chi}\mathrm{d}\chi~\mathrm{d}S|\cos{\chi}|~n_{e,i}^0(r,a,T_e,\chi,\psi),
\end{equation}
with $\beta(T_e)$ the relativistic factor. Hereafter, ionisation electrons are considered to be emitted 
\textit{isotropically}. This assumption may not be accurate given that ionisation electrons should be
produced to some extent along the flow of the general motion of the high-energy electrons of the cascade.
However, since low-energy photons will be emitted isotropically from these non-relativistic electrons,
this assumption does not impact any of the results concerning the incoherent emission of photons
except for a small and irrelevant change of flux of ionisation electrons. 
Under isotropy, the quantity per solid angle unit $n_{e,i}^0(r,a,T_e,\chi,\psi)$ is independent of the 
incidence angles and reduces to $n_{e,i}^0(r,a,T_e)/4\pi$. This yields to the expression of the 
\textit{instantaneous} flux per kinetic energy band, which is the relevant quantity for the following:
\begin{equation}
\label{eqn:fluxelec0}
\phi_{e,i}^0(r,a,T_e)=\frac{c\beta(T_e)f_0(T_e)}{2\left\langle I_0+T_e\right\rangle}~\left\langle\frac{\mathrm{d}E}{\mathrm{d}X}\right\rangle~\rho_m(a)~n_{e,p}(r,a).
\end{equation}

\subsection{Time Evolution of the Flux of Ionisation Electrons}
\label{timeevolution}

The instantaneous flux per kinetic energy band obtained through equation~\ref{eqn:fluxelec0} 
corresponds to the number of secondary electrons per surface, time and kinetic energy units 
left just after the passage of the primary high-energy electrons of the shower. The flux of secondary 
low-energy electrons $\phi_{e,i}(r,a,T_e,t)$, still per kinetic energy band, available at any time $t$ after
the passage of the shower is governed by the interactions that these electrons undergo in the 
atmosphere. In turn, the evolution in time of the function $\phi_{e,i}(r,a,T_e,t)$ can be fully encompassed
in the time dependence of the distribution in kinetic energy $f(T_e,t)$ of the ionisation electrons. 
This evolution is determined by a Boltzmann equation accounting for all the interactions of interest
at work, Boltzmann equation that is now detailed and solved numerically. 

\begin{figure}[!t]
\centering
\includegraphics[width=11cm]{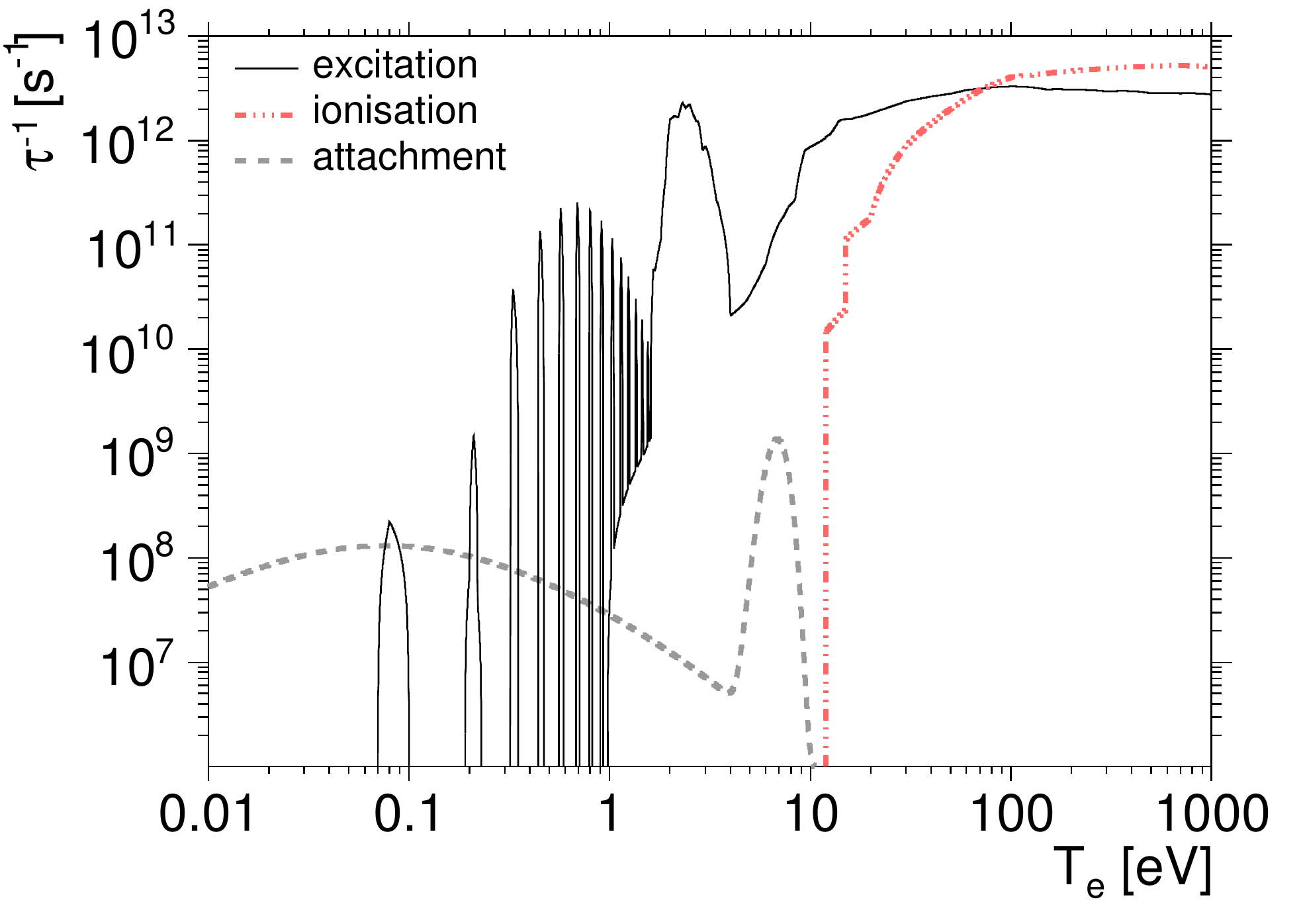}
\caption{\small{Collision rates of interest as a function of electron kinetic energy, at sea level.}}
\label{fig:rates}
\end{figure}

Ionisation electrons can be considered as static in space to a good approximation given their low energy
and given that their rate of disappearance is governed by attachment processes which occur on a time scale 
of at most few hundreds of nanoseconds in the atmospheric layers of interest. It is consequently comfortable 
to neglect the space diffusion term in the Boltzmann equation. The time evolution of the
distribution function $f$ is then exclusively governed by a collision term in the following Boltzmann equation:
\begin{eqnarray}
\label{eqn:boltzmann}
\frac{\partial f}{\partial t}(T_e,t)&=&-n_m(a)c\beta(T_e)\bigg(\sigma_{\mathrm{att}}(T_e)+\sigma_{\mathrm{exc}}(T_e)+\sigma_{\mathrm{ion}}(T_e)\bigg)f(T_e,t) \nonumber \\
&+&n_m(a)c\int_{T_e}^{T_e^{\mathrm{max}}}\mathrm{d}T^\prime_e\beta(T^\prime_e)\bigg(\frac{\mathrm{d}\sigma_{\mathrm{ion}}}{\mathrm{d}T_e}(T^\prime_e, T_e)+\frac{\mathrm{d}\sigma_{\mathrm{ion}}}{\mathrm{d}T_e}(T^\prime_e, T^\prime_e-T_e)\bigg)f(T^\prime_e,t) \nonumber \\
&+&n_m(a)c\int_{T_e}^{T_e^{\mathrm{max}}}\mathrm{d}T^\prime_e\beta(T^\prime_e)\frac{\mathrm{d}\sigma_{\mathrm{exc}}}{\mathrm{d}T_e}(T^\prime_e, T_e)f(T^\prime_e,t),
\end{eqnarray}
where $\sigma_i$ denotes the cross sections of interest (ionisation, excitation of electronic levels and 
attachment processes), and $n_m(a)$ is the density of molecules at an altitude $a$ obtained by multiplying 
the corresponding mass density by the Avogadro number $\mathcal{N}_A$ and by dividing by the 
corresponding molar mass $A$. The first term in the right hand side stands for the disappearance of electrons 
with kinetic energy $T_e$, while the second and third terms stand for the appearance of electrons with kinetic 
energy $T_e$ due to ionisation and excitation (including ro-vibrational excitation) reactions initiated by electrons 
with higher kinetic energy $T^\prime_e$. Note that in the case of ionisation, a second electron emerges from the 
collision with kinetic energy $T^\prime_e-T_e$.

The quantities $\tau^{-1}_i=n_m(a)c\beta(T_e)\sigma_i(T_e)$ characterize the rates of collisions. The electron
attachment processes with oxygen have been studied in detail as a function of the energy, and is here taken
from reference~\cite{Kroll}. The corresponding collision rate leads to a characteristic time scale 
$\tau_{\mathrm{att}}\simeq20~$ns at sea level for energies below 1~eV. The ionisation cross-section is taken 
from reference~\cite{Rapp}, it varies between $\simeq1$ and $\simeq10$ atomic units in the energy range of 
interest, leading to a characteristic time scale between two ionisations below $\sim$ 1 picosecond. Finally, 
collisional rates describing the excitation of the different electronic levels of interest of $O_2$ and $N_2$ 
molecules, including ro-vibrational excitation reactions, are taken from~\cite{Phelps}. Overall, they lead to 
characteristic time scales around $\sim$1 picosecond. The different collision rates are shown in figure~\ref{fig:rates} 
at sea level. For energies above $\simeq 100~$eV, ionisation is the dominant energy loss process, 
while for energies between $\simeq1.7$~eV (corresponding to the threshold to excite one
of the electronic levels of oxygen) and $100~$eV, excitation is the dominant process. Below $\simeq1.7$~eV, 
ro-vibrational excitation reactions compete within some quantised energy ranges with the attachment process which causes a real disappearance of electrons. 
 
\begin{figure}[!t]
\centering
\includegraphics[width=11cm]{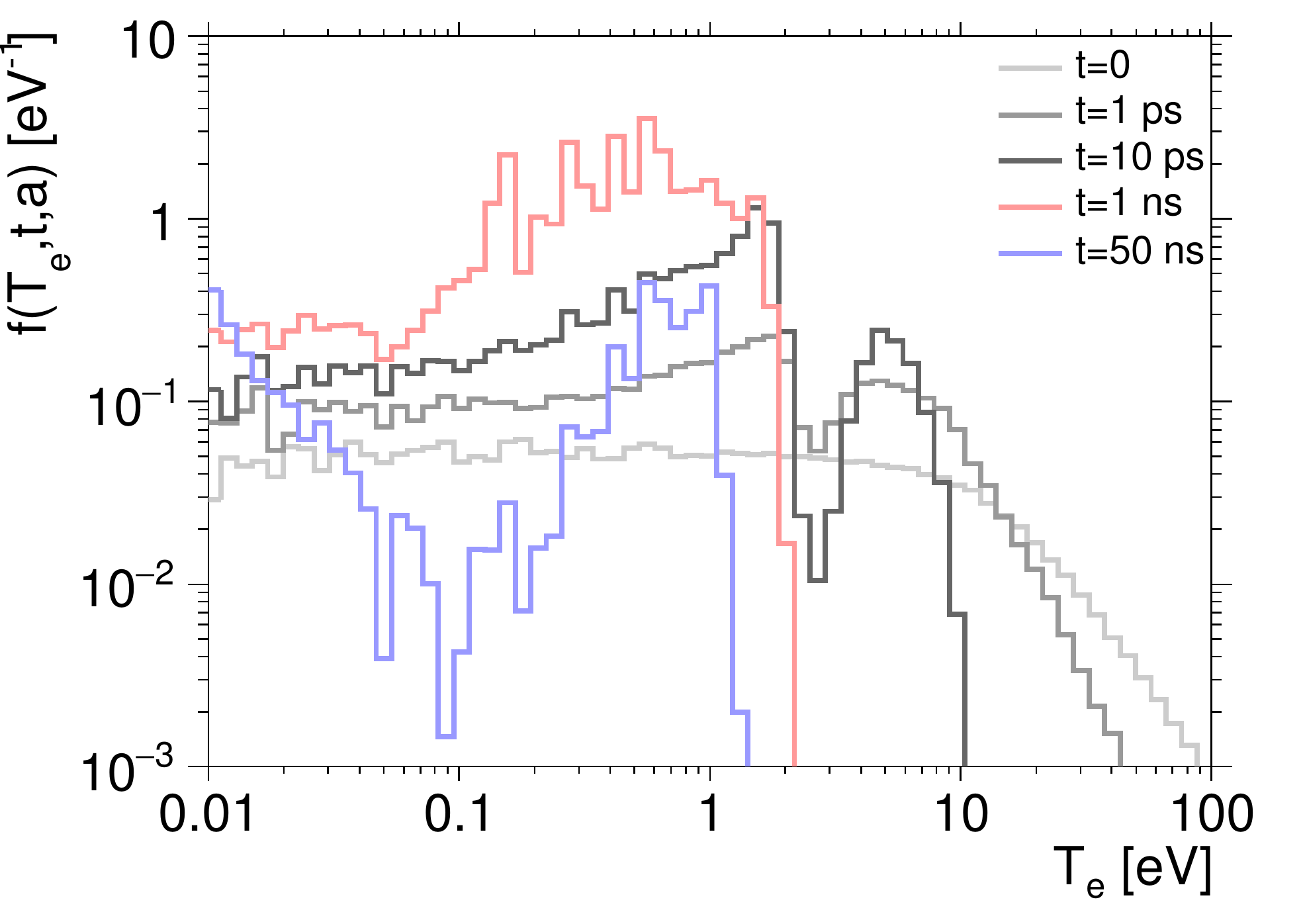}
\caption{\small{Distribution in kinetic energy of the ionisation electrons at different times (at sea level in this example).}}
\label{fig:f}
\end{figure}

The second and third terms in equation~\ref{eqn:boltzmann} account for the migration of electrons with kinetic
energies $T^\prime_e$ to lower energies through ionisation and excitation processes. From the ionisation collision 
rate plotted in figure~\ref{fig:rates}, the population of electrons produced with an energy larger than $\simeq 100~$eV
is expected to be confined in the energy range below $\simeq 100~$eV on a time scale much smaller than 
a nanosecond. New ionisation electrons created through this process undergo the same mechanism. On almost 
the same time scale, the excitation collision rate is expected to confine all electrons in a small energy window 
below $\simeq1.7$~eV. Below this threshold, other excitation processes, mainly ro-vibrational ones, are possible
only within quantised energy ranges. Thus, on a larger time scale (from few nanoseconds to a hundred of nanoseconds), 
the electrons are expected to disappear through the attachment process to oxygen while they shall frequently experience 
reactions modifying their energy spectrum. To quantify precisely this picture, the Boltzmann equation can be solved 
numerically in different ways. We show in figure~\ref{fig:f} the evolution with time of the function $f$ as obtained by 
Monte-Carlo for electrons produced at sea level. After few picoseconds only, the energy distribution $f$ is already 
largely modified by the high collision rate of ionisation for electrons with initial kinetic energies larger than 
$\simeq 100~$eV and by the high collision rate of excitation for electrons with initial kinetic energies smaller than 
$\simeq 100~$eV. After $\simeq 1~$ns, all electron energies get below the lowest excitation threshold belonging
to the continuous spectrum of energy losses. This results in a sharp cutoff in the $f$ function, which confines all electron 
energies below $\simeq 1.7~$eV on a short time scale. On longer time scales, free electrons can still collide through
excitation processes when their energies resonate with the corresponding quantised reactions. This results in a 
slow drift of the sharp cutoff to lower energies as well as a modification of the shape of the $f$ function. Meanwhile, 
electrons disappear through the attachment process. Since the number of electrons is not conserved, the normalisation 
of $f$ against the kinetic energy varies with time; it starts by quickly increasing due to the new electrons produced by 
ionisation, and then decreases in a quasi-exponential way on a longer time scale driven by the attachment process.

\section{Microwave Emission from Molecular Bremsstrahlung: the Free-Free Approach}
\label{mbr}

As long as they remain free, ionisation electrons can produce photons through the process of quasi-elastic collisions 
with neutral molecules in the atmosphere:
\begin{equation}
\label{eqn:mbr}
e^{\pm}+M\rightarrow e^{\pm}+M+\gamma.
\end{equation}
In this approach, the production of photons with energies $h\nu$ corresponds to transitions between unquantised 
energy states of the free electrons ('free-free' transitions). The spectral intensity at ground level can be deduced in 
a straightforward way from the collision rate of ionisation electrons with neutral molecules in air. This approach has 
been shown to be successful for describing, for instance, the production of free-free radiations in collisions 
of low energy electrons with neutral atoms in measurements using drift-tube techniques~\cite{Yamabe}.

By considering the production rate $r_\gamma$ of photons with energy $h\nu$ per volume unit proportional to the target density only, it is then governed by the electron flux and by the free-free cross section $\sigma_{\mathrm{ff}}(T_e,h\nu)$, leading to the following expression:
\begin{equation}
\label{eqn:rate_freefree}
r_{\gamma}(r,a,t,\nu)=n_m(a)~\int_0^{T_e^{\mathrm{max}}} \mathrm{d}T_e~\phi_{e,i}(r,a,T_e,t)~\sigma_{\mathrm{ff}}(T_e,h\nu).
\end{equation}
Possible effects of absorption or suppression of the emission due to destructive interferences of the photons within the interaction zone are neglected at this step. Such effects will be further discussed in section~\ref{attenuation_effects}.

The free-free cross-section as obtained in reference~\cite{Kasyanov} can be related to the electron momentum transfer 
cross-section through:
\begin{equation}
\label{eqn:sigma_ff}
\sigma_{\mathrm{ff}}(T_e,h\nu)=\frac{4}{3\pi}\frac{\alpha^3}{R_y}\left(1-\frac{h\nu}{2T_e}\right)\sqrt{1-\frac{h\nu}{T_e}}T_e\sigma_m(T_e),
\end{equation}
with $\alpha$ the fine-structure constant and $R_y$ the Rydberg constant. For electrons with kinetic energies in the 
range of few tens of eV and photons in the GHz energy range, this expression is very accurately independent of $h\nu$ 
and can be reduced to $\sigma_{\mathrm{ff}}(T_e)=1.211~10^{-8}T_e\sigma_m(T_e)$, with $T_e$ expressed here in units eV. 
The electron momentum transfer cross-section, on the other hand, has been well measured on various targets. Compiled 
tables provided in reference~\cite{Phelps} were used for the following.

As already stressed, the space volume that these electrons can probe during their relatively small lifetime is negligible 
compared to the volume in which an extensive air shower develops. In this way, it is comfortable to consider each electron 
as a point-like source of photons during its whole lifetime. Hence, from the knowledge of the collision rate per volume unit 
$r_\gamma$, the emitted power per volume unit at each point $(r,a)$ can be simply obtained by coupling this rate to the 
energy of the emitted photons, so that the emitted spectral power per volume unit can be written as:
\begin{eqnarray}
\label{eqn:spectral_power_1}
\frac{\mathrm{d}^2P}{\mathrm{d}\nu \mathrm{d}V}(r,a,t)&=&\frac{\mathrm{d}}{\mathrm{d}\nu}\left(h\nu r_\gamma(r,a,t)\right)\\
&=&\frac{hc\rho_m^2(a)\mathcal{N}_A}{2A(I_0+\left\langle T_e\right\rangle)}\left\langle\frac{\mathrm{d}E}{\mathrm{d}X}\right\rangle n_{e,p}(r,a) \tilde{\sigma}(t,a),
\end{eqnarray}
where $\tilde{\sigma}(t,a)$ is an effective cross-section defined as:
\begin{equation}
\label{eqn:sigmatilde}
\tilde{\sigma}(t,a)=\int_0^{T_e^{\mathrm{max}}} \mathrm{d}T_e f(T_e,t,a)\beta(T_e)\sigma_{\mathrm{ff}}(T_e).
\end{equation}

The transparency to photons of the electrons-neutral molecules will be justified in the next section, so that the radiation 
produced by individual electrons-nitrogen/oxygen encounters can be considered here to pass out of the interaction 
volume without absorption or reflection. At any distance $R$, the spectral intensity received from sources contained 
in any infinitesimal volume $\mathrm{d}V$ is proportional to $\mathrm{d}^2P/\mathrm{d}\nu \mathrm{d}V$ times $\mathrm{d}V$ 
and weighted by $4\pi R^2$ given that photons are emitted isotropically\footnote {The assumption on isotropy is justified 
for non-relativistic electrons in the regime where the energy of the photons is low compared to the energy of the 
electrons~\cite{landau}, as this is the case here.} from each source. In this way, the observable spectral intensity at any 
ground position $\mathbf{x}_g$, $\Phi_g$, is simply the sum of the uncorrelated contributions of the individual encounters:
\begin{equation}
\label{eqn:spectral_intensity_0}
\Phi_g(\mathbf{x}_g,t)=\int_0^\infty r\mathrm{d}r\int_0^{2\pi}\mathrm{d}\varphi\int_0^\infty \mathrm{d}a\frac{1}{4\pi R^2(r,\varphi,a)}\frac{\mathrm{d}^2P}{\mathrm{d}\nu \mathrm{d}V}(r,a,t_d(t,r,\varphi,a)).
\end{equation}
Here, $R$ is the distance between the position at ground $\mathbf{x}_g$ and the position of the current source in the 
integration, and $t_d$ is the \textit{delayed} time at which the emission occurred. Fixing the reference time $t_0$ to the 
time at which the shower front crosses the ground, each source at altitude $a$ started emitting radiation at the time the 
shower front passed at that altitude (\textit{i.e.} at $t_0-a/c$). Each photon crossing the ground at time $t$ (with the 
condition that $t\geq t_0$) coming from a source at altitude $a$ and located at a distance $r$ from the shower 
axis was emitted at time $t-Rn(a,\nu)/c$ - with $n(a,\nu)$ the refractive index of the atmosphere integrated along
the line of sight between the emission point and the observer. The delayed time $t_d(t,r,\varphi,a,\nu)$, denoted 
hereafter $t_d$ only for convenience, is thus expressed as:
\begin{equation}
\label{eqn:delay}
t_d\equiv t_d(t,r,\varphi,a,\nu)=t-t_0-\left(\frac{R(r,\varphi,a)n(a,\nu)}{c}-\frac{a}{c}\right).
\end{equation}
With the evident condition that emissions occur only at $t_d\geq 0$, a Heaviside function denoted $\Theta$ is introduced leading to the following semi-analytical expression for the observable spectral intensity:
\begin{equation}
\label{eqn:spectral_intensity_1}
\Phi_g(\mathbf{x}_g,t)=\frac{hc\mathcal{N}_A}{8\pi A(I_0+\left\langle T_e\right\rangle)}\left\langle\frac{\mathrm{d}E}{\mathrm{d}X}\right\rangle\int_0^\infty r\mathrm{d}r\int_0^{2\pi}\mathrm{d}\varphi\int_0^\infty \mathrm{d}a~\frac{\rho_m^2(a)n_{e,p}(r,a)}{R^2(r,\varphi,a)}~\tilde{\sigma}({t_d,a})~\Theta(t_d).
\end{equation}

Besides the free-free approach presented here, an independent estimation of the radiated power using the 
classical field theory formalism can be applied, resulting in the same predictions. Detailed expressions in this 
frame can be found in the Appendix. 

\section{Possible Attenuation Effects in Molecular Bremsstrahlung Radiation}
\label{attenuation_effects}

\subsection{Absorption Effects}
\label{absorption}

In addition to free-free emissions, ionisation electrons can also experience inverse Brems\-strah\-lung and stimulated
Bremsstrahlung within the electrons-neutral molecules plasma. A convenient way to quantify the size of these effects 
is to calculate the absorption coefficient, $\alpha_\nu$, defined as the relative attenuation per unit length of the emitted
photons. The absorption coefficient is defined as the net balance between the number of absorbed photons per unit
length subtracted to the number of stimulated emitted photons (due to a photon that causes an electron in the potential 
of a neutral molecule to emit another photon of the same frequency) per unit length. 

To derive the absorption coefficient, it is convenient to introduce the emitted spectral power per volume unit \textit{at a 
fixed energy} $T_e$, denoted $\eta_{\nu,V}(T_e,r,a)$, in the same way as in reference~\cite{Bekefi} for instance. 
Then, equation~\ref{eqn:spectral_power_1} can be re-written as:
\begin{equation}
\label{eqn:spectral_power_2}
\frac{\mathrm{d}^2P}{\mathrm{d}\nu \mathrm{d}V}(r,a,t)=\int \mathrm{d}T_e f(T_e,t,a) \eta_{\nu,V}(T_e,r,a).
\end{equation}
The absorption coefficient $\alpha_\nu$ is then known to be related to $\eta_{\nu,V}$ through~\cite{Bekefi}:
\begin{equation}
\label{eqn:alphanu_1}
\alpha_\nu=\frac{c^2}{h\nu^3}\int \mathrm{d}T_e^\prime \bigg[f_0(T_e,a)-f_0(T_e^\prime,a)\bigg] \eta_{\nu,V}(T_e^\prime,r,a).
\end{equation}
This expression accounts for both the absorption of a photon of energy $h\nu$ by an electron with initial energy $T_e$
and a final one $T_e^\prime$ and the stimulated emission due to a photon that causes a neighboring electron to emit
another photon of the same energy within the electrons-neutral molecules plasma. Given the low frequencies of the 
photons considered here compared to the mean energy of the electrons, expanding $f_0(T_e,a)$ to first order in energy
leads to:
\begin{equation}
f_0(T_e^\prime,a)=f_0(T_e,a)+h\nu\frac{\partial f_0}{\partial T_e^\prime},
\end{equation}
so that the absorption coefficient $\alpha_\nu$ reads as:
\begin{equation}
\alpha_\nu(r,a)=-\frac{c^2}{\nu^2}\int \mathrm{d}T_e \frac{\partial f_0(T_e,a)}{\partial T_e} \eta_{\nu,V}(T_e,r,a) .
\end{equation}
Injecting explicitly the expression of $\eta_{\nu,V}$ into this expression, $\alpha_\nu$ turns out to read:
\begin{equation}
\label{eqn:absorption}
\alpha_\nu(r,a)=-\frac{hc^3 \rho^2(a)\mathcal{N}_A}{2A(I_0+\left\langle T_e\right\rangle)}\frac{1}{\nu^2}\left\langle\frac{\mathrm{d}E}{\mathrm{d}X}\right\rangle n_{e,p}(r,a)\int \mathrm{d}T_e \beta(T_e)\sigma_{\mathrm{ff}}(T_e)\frac{\partial f_0(T_e,a)}{\partial T_e}.
\end{equation}
Close to the shower core and to the maximum of shower development (that is, within the denser 
plasma region), this leads to $\nu^2\alpha_\nu\simeq 10^{-4}$~m$^{-1}$~Hz$^2$. At GHz frequencies, 
the absorption is thus negligible. 

\subsection{Suppression Effects}
\label{suppression}

The spectral intensity predicted by equation~\ref{eqn:spectral_intensity_1} is based on the assumption
that the emitted radiation passes out of the interaction volume without undergoing any dispersive properties
of the plasma caused by the successive interactions of the electrons.  
 
Dispersive properties are commonly described on a macroscopic basis by a dielectric coefficient. This coefficient 
allows a derivation of the absorption coefficient which, in contrast to the previous derivation, accounts 
for successive collisions within the radiation formation zone of each electron-neutral collision. The
effect of the successive collisions, known as the plasma dispersion effect~\cite{Bekefi}, can lead to 
destructive interferences of the radiated fields. 

Accounting for the coupling of electrons to the emitted radiation turns out to be a difficult task
here, since it consists in considering an additional term in equation~\ref{eqn:boltzmann} proportional to 
$\partial f/\partial T_e$ times the radiated electric fields. To get an order of magnitude 
of the effects which might be at work, we adopt the commonly used method consisting in linearising 
a simplified Boltzmann equation pertaining only to the case of a distribution function $f(T_e)$
stationary in time in the absence of any emitted radiation. In this case, it can be shown that the 
plasma dispersion effects result in a suppression factor in the integrand of $\tilde{\sigma}$ 
defined in equation~\ref{eqn:sigmatilde} reading as~\cite{Bekefi}:
\begin{equation}
\frac{1}{1+(\nu_c(T_e,t)/\nu)^2},
\end{equation}
where $\nu_c(T_e,t)$ is the time-dependent rate of inelastic collisions of the electrons of kinetic energy $T_e$. 
From the analysis of section~\ref{timeevolution}, this collision rate amounts to several THz within the
first nanosecond for highly energetic electrons and then decreases to the level of a few tens of MHz.
Consequently, for frequencies $\nu$ around the GHz, the suppression factor can be important only 
during the first nanosecond, as long as the collision rate is much larger than the frequency considered
for the radiation field.

It is to be noted that the suppression factor aforementioned is derived by means of
a series of hypotheses which are not really relevant in the case considered here. However, it is clear
that such plasma dispersion effects can be important only when $\nu_c(T_e,t)$ is larger than $\nu$. As
a kind of proxy to probe these effects in the most pessimistic way, the impact can be evaluated by 
suppressing the emission as long as $\nu_c(T_e,t)>\nu$.

\section{Discussion}
\label{discussion}

Although semi-analytical integrations of equation~\ref{eqn:spectral_intensity_1} are possible
without random number generators, a Monte-Carlo sampling of the integrand function
in $r$ and $\varphi$ allows a much faster integration in terms of CPU time. Results
presented here have thus been obtained using random number generators to carry out 
these particular integrations.

\begin{figure}[!t]
\centering
\includegraphics[width=11cm]{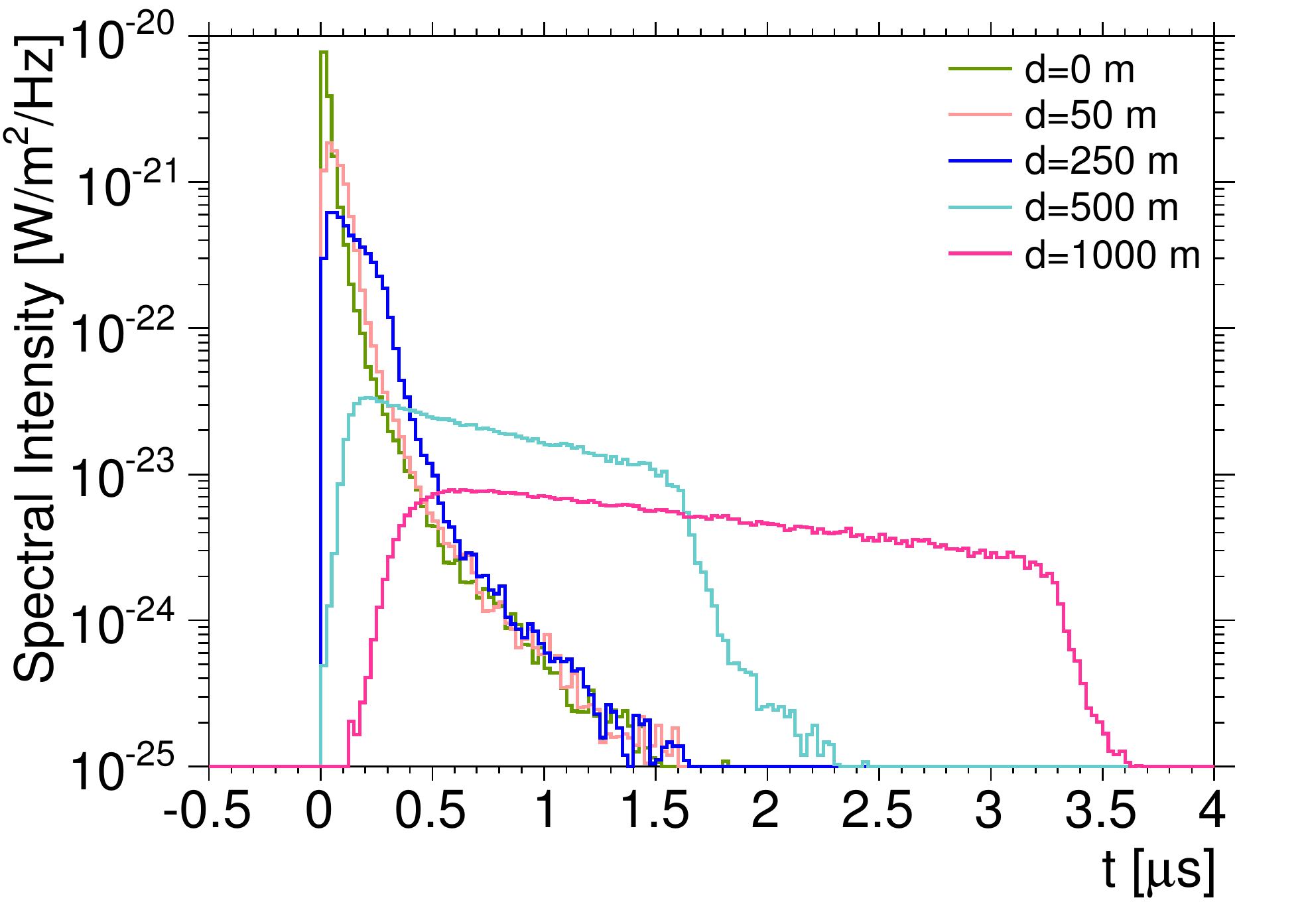}
\caption{\small{ Spectral intensity as a function of time expected at different distances
from the shower core at ground level, for a vertical shower with energy $10^{17.5}$~eV.}}
\label{fig:all}
\end{figure}

\begin{figure}[!t]
\centering
\includegraphics[width=11cm]{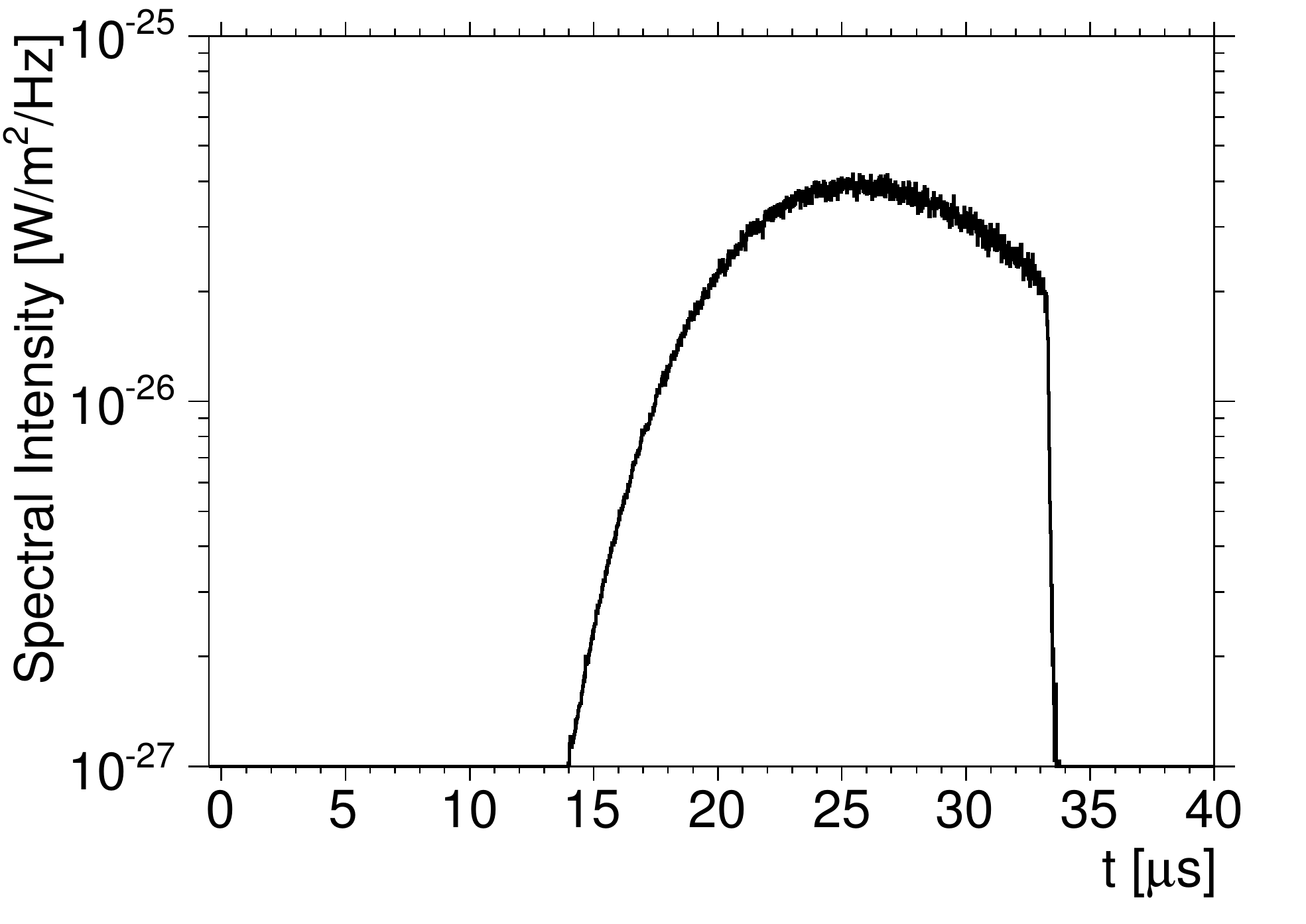}
\caption{\small{\small{Spectral intensity as a function of time expected at 10~km
from the shower core at ground level, for a vertical shower with energy $10^{17.5}$~eV.}}}
\label{fig:10km}
\end{figure}

A vertical proton shower of $10^{17.5}$~eV is used as a proxy to illustrate the estimation of the spectral 
intensity expected from molecular Bremsstrahlung radiation presented in section~\ref{mbr}. The parameterisation
of the atmosphere selected here is the widely used \textit{US standard atmosphere}, based on experimental 
data~\cite{NASA-atm}. The parameterisation of the wavelength dependence of the refractive index is taken
from~\cite{Weast}. 

Experimental setups using regularly spaced antennas oriented vertically or nearly vertically detect showers 
crossing the field-of-view and impacting the ground. The spectral intensity expected at different distances from 
the shower core at ground level as derived from equation~\ref{eqn:spectral_intensity_1} is shown in 
figure~\ref{fig:all}. This figure is relevant for the case of the currently running EASIER installation at the Pierre 
Auger Observatory~\cite{Gaior} or for the CROME experiment~\cite{CROME}. 
It is seen that the spectral intensity is rapidly decreasing in amplitude for increasing distances to the shower core. 
The duration of the signals at different distances can be understood from the characteristic time scale of the
attachment process on the one hand, and from the different regions of the shower that can be probed for different
positions at ground on the other hand. 
This result is in disagreement with expectations found in~\cite{KITicrc2013}, in which the signal time profiles at these distances are difficult to interpret when accounting for all-interaction cross sections. Furthermore, the reported short signal duration and the low amplitudes are not found to be reproducible by considering only the attachment process as in~\cite{KITicrc2013}, .

An alternative detection method of microwave radiation is the use of large aperture receivers pointing just above 
the horizon to observe the longitudinal profiles of the showers at large distances, such as MIDAS and 
AMBER installations at the Pierre Auger observatory~\cite{Gaior,MIDAS}. The received power as a function of time 
at a distance of 10~km from the shower axis is shown in figure~\ref{fig:10km}. It turns out to be $\simeq 4.0~10^{-26}$~W~m$^{-2}$~Hz$^{-1}$. To probe the maximal impact of the plasma dispersion effects as discussed in the previous section, the same simulation 
is repeated with the condition of suppressing the emission as long as $\nu_c(T_e,t)>\nu$. At 10~km from the shower axis, 
the signal is found to be $\simeq 3.8~10^{-26}$~W~m$^{-2}$~Hz$^{-1}$, not significantly different from the spectral intensity 
value obtained without accounting for any suppression effect. This gives an idea about the small systematic uncertainty 
which affects the estimate due to possible collective suppression within the plasma.

The estimated intensity is found to be smaller than the one reported in reference~\cite{Gorham} by a factor $\simeq 70$ when scaling the beam measurements to air showers. Recent results from~\cite{Conti} reported a measurement of an anisotropic distribution of the molecular Brems\-strah\-lung radiation. The interpretation of this result in the frame of air shower physics is not yet clear and needs further investigation. We note that results from laboratory measurements are still to be confirmed by independent experiments~\cite{MAYBE,AMY}. 

In air shower experiments, an important difficulty arises from the separation power between the molecular Brems\-strah\-lung radiation signals and signals due to the geomagnetic radiation. Close to the shower core, the measured signal amplitude from geomagnetic radiation~\cite{CROME} is greater than or of the same order as the one expected from molecular Brems\-strah\-lung radiation found in this study. However, the expected signal duration appears significantly larger (about a factor 100) comparing to the measured signal from geomagnetic radiation. Together with the unpolarised nature of the signal, there are thus possible
experimental signatures for the molecular Bremsstrahlung radiation which would allow its identification, provided that the experimental setup is sensitive enough.
Based on this study, significant increases in sensitivity should thus be achieved from an experimental point of view to be able to detect showers induced by ultra-high energy cosmic rays by means of the molecular Bremsstrahlung emission mechanism.

\section*{Acknowledgements}

We acknowledge the support of the French Agence Nationale de la Recherche (ANR) under reference ANR-12-BS05-0005-01. 
We thank Auger members who have participated in the review of this paper, in particular the EASIER group. We also thank Jaime 
Rosado Velez for his valuable comments on the estimation of the mean number of ionization electrons and Felix Werner for his 
careful reading of the document.

\newpage
\begin{center}
      {\bf APPENDIX}
    \end{center}
\section*  {Classical Field theory approach in Molecular Bremsstrahlung Radiation estimation}
 \label{cl}

From the point of view of the classical field theory, the radiated power by the
ionisation electrons is associated to the deviations caused by the collisions
with the neutral molecules. In this framework, although the formal expression of 
the spectral intensity expected at ground level is unchanged with respect to 
equation~\ref{eqn:spectral_intensity_0}, the expression of the emitted spectral power 
per volume unit has to be revised to remove any reference to free-free transitions.

\begin{figure}[!h]
\centering
\includegraphics[width=8cm]{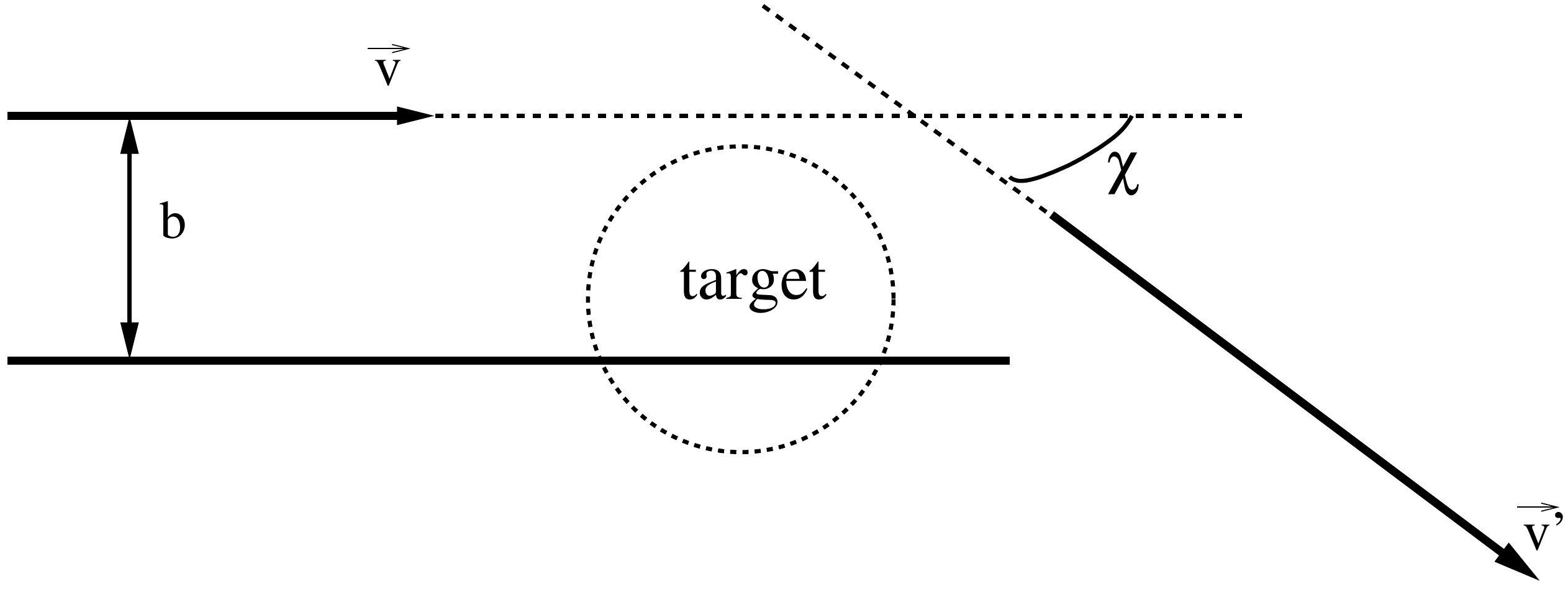}
\caption{\small{Geometry of a classical binary collision. The vector change in velocity is $|\Delta \mathbf{v}|=2v\sin{(\chi/2)}$.}}
\label{fig:collision}
\end{figure}

In a classical way, when an electron approaches a neutral molecule, the electric
field of the electron \textit{polarises} the neutral molecule. This polarisation
gives rise to a dipole moment which induces an attractive interaction potential at a
short distance range~: $V(d)\propto d^{-4}$ - with $d$ the distance between the 
electron and the molecule. The time-dependent radiated power during the interaction 
is known to obey the Larmor formula~: 
$p_e(t)=e^2\left|\dot{\mathbf{v}}(t)\right|^2/6\pi\epsilon_0 c^3$, with $e$ the 
elementary charge and $\epsilon_0$ the vacuum permittivity. Then, by making use of the 
Parseval identity, one can derive directly the frequency spectrum of the radiated 
energy for one collision as:
\begin{equation}
\frac{\mathrm{d}E}{\mathrm{d}\nu}(\nu,v,\chi)=\frac{e^2}{3\pi\epsilon_0 c^3}\left|\int_{-\infty}^\infty \mathrm{d}t~\dot{\mathbf{v}}(t)\exp{(-i2\pi\nu t)}\right|^2.
\end{equation}
Since the interaction potential acts at a short distance range only, the average
time during which the interaction is taking place can be estimated as $\Delta t\simeq b/\left<v\right>$,
with $b$ the impact parameter of the collision. For typical electron velocities of
the order of a few percents of the speed of light, a realistic order of magnitude for
$\Delta t$ is $\simeq 10^{-16}~$s so that for GHz frequencies, $2\pi\nu \tau\ll 1$ for any $\tau$ within $\Delta t$ . Hence, 
the argument in the exponential of the integrand is very small during the time where the electron
is accelerated, so that for an electron undergoing a total deviation by an angle $\chi$, 
the corresponding frequency spectrum of the radiated energy can be accurately estimated as:
\begin{eqnarray}
\frac{\mathrm{d}E}{\mathrm{d}\nu}(v,\chi)&=&\frac{e^2}{3\pi\epsilon_0 c^3}\left|\Delta \mathbf{v}\right|^2\\
&=&\frac{2e^2}{3\pi\epsilon_0 c^3}v^2(1-\cos{\chi}),
\end{eqnarray}
where the expression of $\left|\Delta \mathbf{v}\right|$ has been easily derived from
the geometry of the collision depicted in figure~\ref{fig:collision}.

The collision rate per volume unit, per electron velocity band and per solid angle unit
(with here $d\Omega=\sin{\chi}d\chi d\psi$) is governed by the same ingredients as in the
previous section, except that the free-free cross-section is now replaced by the classical 
differential cross-section:
\begin{equation}
\frac{\mathrm{d}^2r_{\gamma}^{\mathrm{cl}}}{\mathrm{d}v\mathrm{d}\Omega}(r,a,t,v,\chi)=n_m(a)~\phi_{e,i}(r,a,v,t)~\frac{\mathrm{d}\sigma^{\mathrm{cl}}}{\mathrm{d}\Omega}(v,\chi),
\end{equation}
where the flux of ionisation electrons is now expressed, for convenience, per velocity band
instead of kinetic energy band. This is achieved by expressing the kinetic energy in terms
of the velocity and by substituting the normalised $f$ function by the one obtained through
the relevant Jacobian transformation:
\begin{equation}
\tilde{f}(v,t)=\frac{m_ev}{(1-(v/c)^2)^{3/2}}f(T_e(v),t).
\end{equation}

Since the collision rate is independent of the frequency, the emitted spectral power per volume 
unit can be obtained by integrating this collision rate directly coupled to the frequency spectrum 
of the energy radiated per collision over solid angle and the velocity $v$. This leads to:
\begin{equation}
\label{eqn:spectral_power_2}
\frac{\mathrm{d}^2P^{\mathrm{cl}}}{\mathrm{d}\nu \mathrm{d}V}(r,a,t)=\frac{e^2\rho_m^2(a)\mathcal{N}_A}{3\pi\epsilon_0A(I_0+\left\langle T_e\right\rangle)}\left\langle\frac{\mathrm{d}E}{\mathrm{d}X}\right\rangle~n_{e,p}(r,a)~\tilde{\sigma}^{\mathrm{cl}}(t,a),
\end{equation}
with the classical effective cross-section defined as:
\begin{equation}
\label{eqn:sigmatildecl}
\tilde{\sigma}^{\mathrm{cl}}(t,a)=\int \mathrm{d}v~\tilde{f}(v,t)\left(\frac{v}{c}\right)^3\int \mathrm{d}\Omega~(1-\cos{\chi})\frac{\mathrm{d}\sigma^{\mathrm{cl}}}{\mathrm{d}\Omega}(v,\chi).
\end{equation}
Note that the result of the solid angle integration is, by definition, the 
momentum transfer cross-section $\sigma_m(T_e(v))$. 

Hence, it is clear by identification that equation~\ref{eqn:spectral_power_2} can be equivalent to 
equation~\ref{eqn:spectral_power_1} only if the following correspondence holds~:
\begin{equation}
hc\tilde{\sigma}(t,a) \rightarrow \frac{2e^2}{3\pi\epsilon_0}\tilde{\sigma}^{\mathrm{cl}}(t,a). 
\end{equation}
And, it turns out that both energy volumes equal, for nitrogen targets, $t=1~$ns and at sea level for instance, 
to $\simeq 4.5~10^{-37}$~eV~m$^3$. Given the low energy of the photons considered here compared to the 
electron energies, the classical approach results in the same prediction as the free-free approach.

\newpage
\section*{References}

\end{document}